# Development of readout electronics a novel beam monitoring system for ion research facility accelerator

Honglin Zhang [a,b], Haibo Yang [a,b], Xianqin Li [a], Yuezhao Zhang [a], Xiangming Sun [c], Dong Wang [c], Ran Huang [a,b], Yao Wang [a], Wei Zhou [a], Chengxin Zhao [a,b*]

[a] Institute of Modern Physics, Chinese Academy of Sciences, Lanzhou, 730000, China

[b] School of Nuclear Science and Technology, University of Chinese Academy of Sciences, Beijing, 100049, China

[c] Central China Normal University, Wuhan, 430079, China

**Abstract:** This article presents the readout electronics of a novel beam monitoring system for ion research facility accelerator. The readout electronics are divided into Front-end Card (FEC) and Readout Control Unit (RCU). FEC uses Topmetal-II$^-$ to processes the energy of the hitting particles and convert it into a voltage signal. The main function of RCU is to digitize the analog output signal of FEC and format the raw data. On the other hand, the RCU also processes the control commands from the host and distributes the commands according to the mapping. The readout electronic has been characterized and calibrated in the laboratory, and have been installed with the detector. Implementation and testing of readout electronics have been discussed.

## 1 Introduction

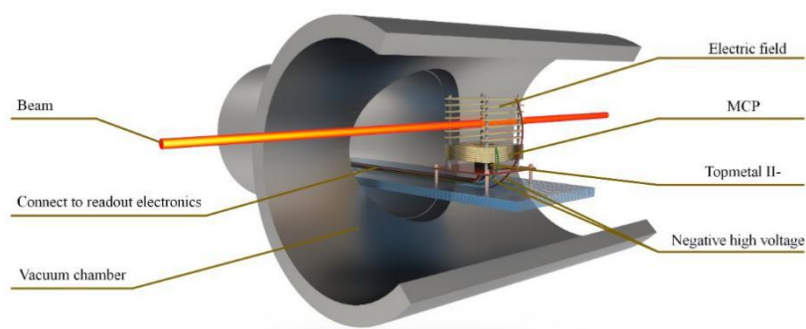

Figure 1. Design prototype of a new beam monitoring system

Several heavy-ion research facilities are hosted by the Institute of Modern Physics, Chinese Academy of Sciences at Lanzhou, China, such as the 320 kV highly charged ion beam platform [1] (the 320kV platform) and the Cooling Storage Ring of the Heavy Ion Research Facility in Lanzhou (HIRFL-CSR) [2][3]. These research facilities together provide various ion beams to study physics and related applications. To monitor the status of the beam in these accelerators, beam profile monitoring system has been designed. In the 320kV platform, the faraday cup array [?] is adopted for beam profile monitoring. In the HIRLF-CSR, a beam position monitor [4] is used. These beam profile monitor has shown stale and well performance. However, there are two major points that need to be improved. Firstly, the spatial resolution of these monitors is only in the scale of several hundred μm [4]. Furthermore, the monitors need to interact with the beam which lies difficulties on real time monitoring. Therefore, a novel Residue Gas Ion Beam Profile

Monitor (RGIPM), which can monitor the beam in a non-blocking manner with the spatial resolution of a few tens of μm, has been designed [5].

Figure 1 shows the typology of the RGIPM. The residual gas in the tube of the accelerator (ultra-high vacuum condition) is ionized by the heavy-ion beam. The electrons generated during the ionization is multiplicated by the Micro-channel Plate (MCP). Driven by the electric field, the electron clusters are then collected by a novel silicon pixel detector – the Topmetal II- [6][7]. Beam is not interfered by the RGIPM, that is, the profile of the beam can be monitored in a real-time manner. Charge collected by the Topmetal II- goes into the readout system.

## 2 The Readout Electronics

As shown in Figure 2, the readout system consists of the Front-End Card (FEC) and the Readout Control Unit (RCU). The FEC measures the energy of the electrons that hit on the detectors and converts it into electric signal. The RCU processes the data from the FEC and transmits it to the upper-stream monitoring system via 1 Giga-bit Ethernet Interface. Each RCU can control up to 8 FECs simultaneously.

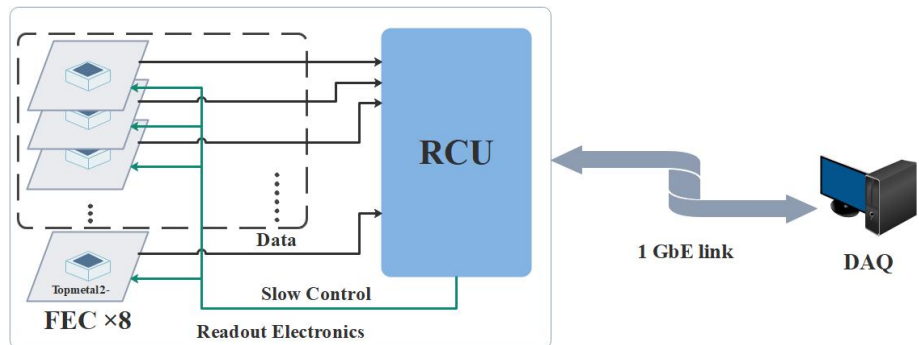

Figure 2. Prototype diagram of readout electronics

### 2.1 The Front-end Card

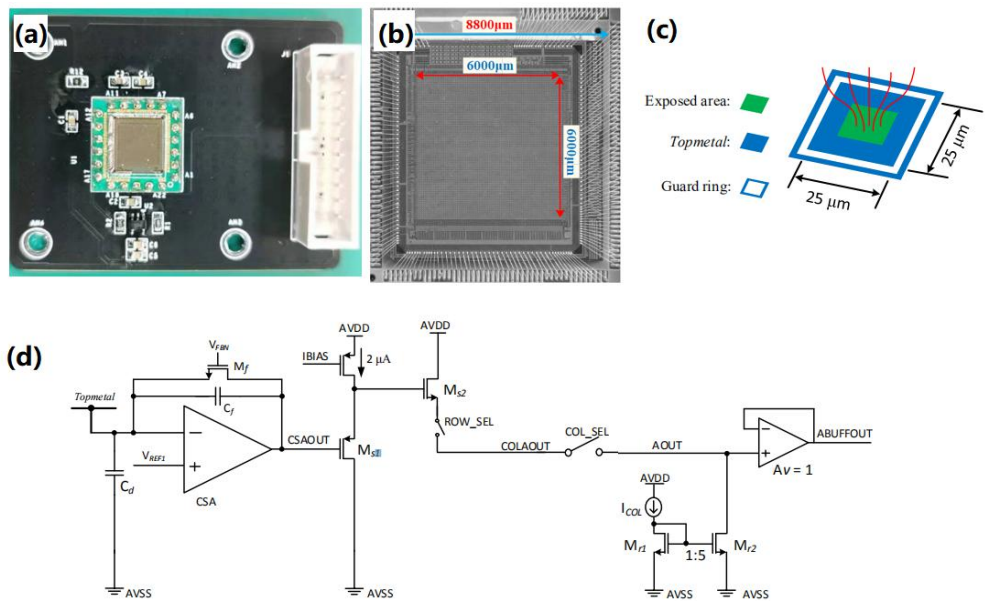

Figure 3. (a) Picture of the FEC (b) the Topmetal II- pixel chip [6] (c) The Topmetal Pad (d) signal path in the Topmetal II-

Figure 3 (a) shows the picture of the front-end card, on which the main component is the Topmetal-II- pixel detector (Figure 3(b)) [6]. Each Topmetal-II- ASIC has 72×72 pixels with the size of 83um×83um. In each pixel, charge is collected by the metal plate -Topmetal (blue plate in Figure 3(c)), which is located at the right-bottom corner of each pixel. The size of the Topmetal plate is 25 μm × 25 μm and the size of exposed area in the center of the Topmetal is 15 μm × 15 μm (green plate in figure 3(c)). The collected charge goes directly into the readout circuity in the ASIC (Figure 3(d)). No detector leakage current occurs in the ASIC since the Topmetal collects charge directly from the surround media. In the readout circuity, the charge is converted into voltage signal by the low-noise Charge Sensitive Amplifier (CSA), then transmits to the Readout Control Unit through through two source-follower stages (Ms1 and Ms2 in Figure 3 (d)) and an analog output buffer. All the 5184 pixels in the Topmetal II- ASIC are read out in rolling-shutter scheme, that is, from 1st pixel to 5184th pixel. The selection on each individual pixel is realized by the column-selection switch and the row-selection switch. With this analog output method, full waveform of the charged collected by the Topmetal II- detector can be accessed. Power, clock and control signals of the Topmetal-II- ASIC are provided by the RCU.

## 2.2 The Readout Control Unit

As show in Figure 4, the RCU mainly consists of the main FPGA (Xilinx Kintex-770T), the shaping circuit, the ADC(AD9252 [13]), two 32bit-DDR3 memories and the Ethernet Interface.

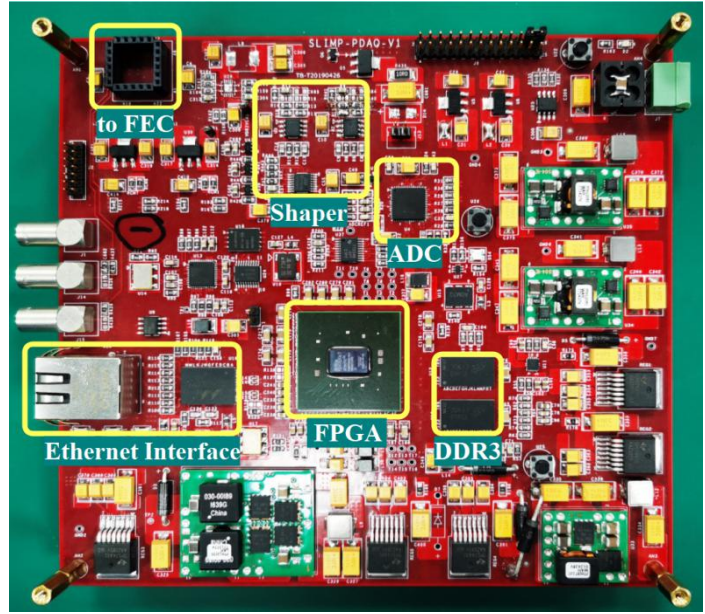

Figure 4. First prototype of the RCU

### 2.2.1 The shaping circuit

As shown in Figure 5, the shaping circuit is in single-ended to differential mode, which is constituted of a single-ended to differential operational amplifier (THS4521 [15]) and a DAC (DAC8568 [14]). The operational amplifier converts the single-ended signal from the FEC into differential signal. Gain (A) of operational amplifier is a function of $R_f$ and $R_g$ as in Equation 1.

$$A = \frac{R_f}{R_g} \quad \text{(Equation 1)}$$

Bias voltage of the operational amplifier is provided by the DAC, whose output can be tuned to adjusts the baseline of the operational amplifier output and improve the signal to noise ratio of the

ADC input.

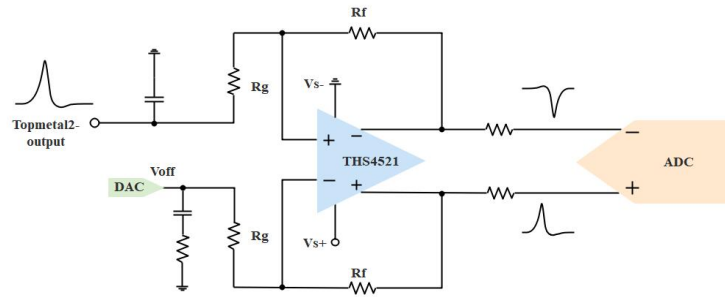

Figure 5. The shaping circuits

### 2.2.2 Firmware design

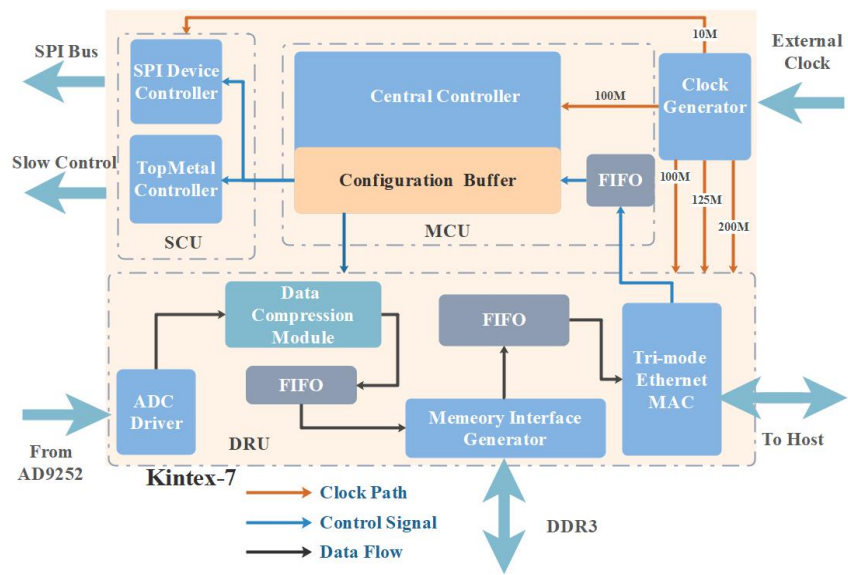

Figure 6. Block diagram of firmware design

Figure 6 shows the firmware design in the FPGA, which consists of the Control Unit and the Readout Unit. In the Control Unit, the Central Controller receives control commands from the upper-stream computer through the the Tri-mode Ethernet MAC [11], stores them in the configuration buffer, and decodes the commands. Upon these commands, the Central Controller controls the SPI Device Controller, the Topmetal Controller and the Readout Unit. The SPI Device Controller configures the on-board ADC and DAC. The Topmetal Controller configures the Topmetal-II- ASIC. In the Readout Unit, the serial data from the ADC is aligned and converted into parallel data by the ADC driver. The data is compressed by the Data Compression Module in a lossy scheme. The compressed data is stored into the on-board DDR3 memories by the Memory Interface Generator[15]. Once the shipping commands comes from the Central Controller, the Memory Interface Generator reads the data from the DDR3 memory and ships it to the the data computer through the Tri-mode Ethernet MAC. Thus, the First-In-First-Out (FIFO) buffer is placed at between the Data Compression Module and Readout Controller to avoid the effects of Clock Domain Crossing. The Tri-mode Ethernet MAC and Memory Interface Generator are based on IP core from the Xilinx Cooperation.

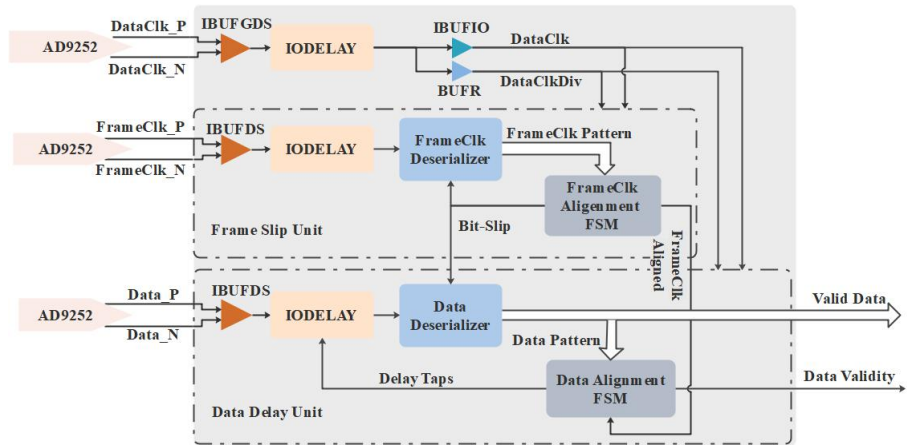

Figure 7. Simplified schematic of ADC driver

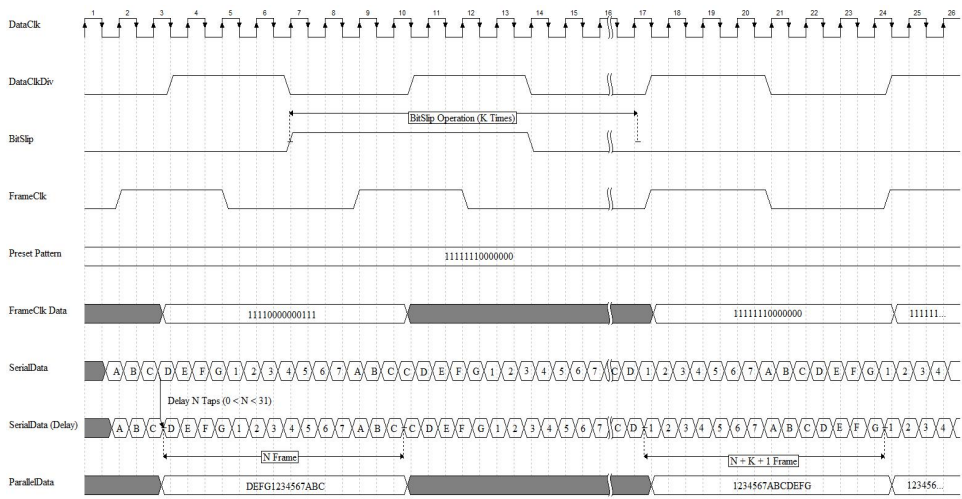

Figure 8. Delay and bit-slip operation timing diagram

The sampling rate of the on-board ADC is up to 50Msps. Thus the alignment procedure between the clock and data has been carefully designed, as shown in Figure 7. Figure 8 shows the timing diagram of the ADC data readout. Each data frame from the on-board ADC consists of 14-bit serial data. The data bits are captured at data clock rate into an input serial-to-parallel register in the Data Deserializer. The data is then transferred from the serial-to-parallel to the parallel register at the DataClkDiv clock, which is 7 divided data clock. Each frame of ADC data is identified by the frame clock. The the frame clock is up to 50MHz and the data clock is up to 350MHz, thus the alignment procedure has been carefully designed, as shown in Figure 7. Firstly, the Frame Alignment Unit aligns the data clock and the DataClkDiv. The frame clock is captured as data by the data clock and compares with the target pattern as of "11111110000000". Bit-slip. operation, which delays the frame clock by one data clock cycle, will be performed unit the target pattern is obtained. This ensures the serial-to-parallel conversion on each data frame are performed in a correct manner. After which, the Data Alignment Unit handles the shift between each data bit and the edge of the data clock, which is caused by the delay on the transmission line. The delay on the IO components [9] in the FPGA can be configured up to 32 taps, each of which is 78ps with the reference clock at 200MHz. In the initialization process, the ADC will be set in test mode and generate certain data pattern. The Data Alignment FSM compares the received data with the target pattern. The number of taps will be increased in the steps of one until the data pattern is constantly obtained. The whole alignment process can be performed at anytime upon the reset signal from upper-stream computer.

## 3 Performance characterization

Performance of the readout electronics have been characterized in the lab. Stability test of the data and clock alignment from the ADC has been performed. In the Xilinx Kintex-7-70T, the IDELAYE2 (IDELAYE2 is an IO block component in Xilinx's 7 series FPGA.) delay line (IO delay) on the ADC data can be configured up to 32 taps, each of which is 78ps, while the reference clock of the IDELAYE2 component is set as 200 MHz. A 10 kHz ramp signal generated by a precise pulse generator (Tekronix AFG3252C) is used to provide input to the RCU. The delay value of the IDELAYE2 component has been set from 4 taps to 15 taps (312ps to 1170ps) to match the delay of the data from the ADC into the FPGA. Table 1 shows the data validity with each delay taps, where the for each tap with valid data, the test setup has been running for ~4 hours. This proves the alignment of the ADC data and clock output has a stable working region. Figure 9 shows the waveform of the data with well alignment and invalid alignment.

*Table 1. Data performance with different number of taps*

| Number of Taps | Data validity | Number of Taps | Data validity |
|---|---|---|---|
| 4 | N | 10 | Y |
| 5 | N | 11 | Y |
| 6 | N | 12 | Y |
| 7 | Y | 13 | Y |
| 8 | Y | 14 | Y |
| 9 | Y | 15 | N |

*Note: "Y" means passed the IODELAY test, and "N" means failed.*

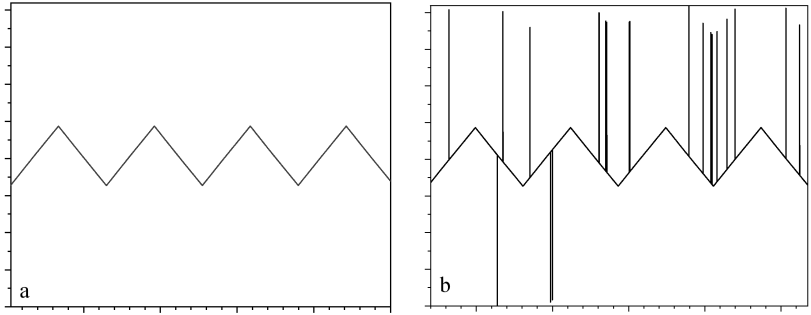

Figure 9. (a) Data with well alignment (b) Data with invalid alignment

Linearity of the RCU and the whole readout system has been characterized in the lab. In the former test, a precise pulse generator (Tekronix AFG3252C) has been used to provide the input to the RCU. The injected signal goes through the whole signal chain on the RCU (shaper, ADC, FPGA, DDR memory and Ethernet Interface) then transmits to the data computer. The input signal varies from -1.5V to 1.5V in the steps of 0.1V. The readout system has been running for 48 hours and the tests have been performed every 8 hours. Based on the recorded statistics, linear fitting function between each dedicated input value and the average output of the RCU has been derived. For each input value, the non-linearity is calculated as Equation 2, where $BinOut_{test_n}$ is the output measured in the tests and $BinOut_{fit}$ is the output calculated with the fitting function.

$$System\ variation(input) = (\underset{-1.5 << n << 1.5}{Max}\{|BinOut_{test_n} - BinOut_{fit}|\}) \quad (Equation\ 2)$$

Figure 10 (a) shows the output of the RCU as a linear fitting function of the input. Figure 10 (b) shows the nonlinearity of each certain input value, where the maximum variation is ~0.6%.

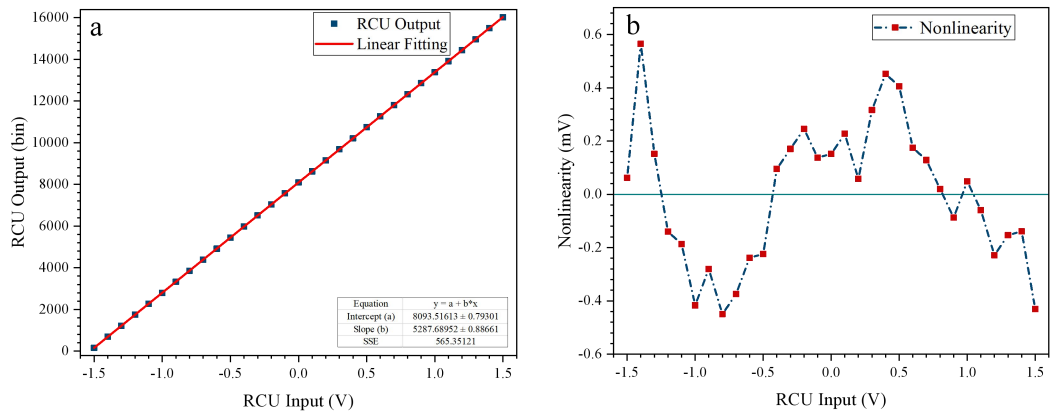

Figure 10. (a) Output as function of input (b) Nonlinearity of each input

Linearity of the whole readout system, including the FEC and the RCU has also be characterized in the lab. Through giving certain voltage to the guarding ring of each pixel in the Topmetal II- ASIC chip, charge can be directly injected into charge sensitive amplifier in each pixel then go through the whole readout chain. In our test, the input voltage varies from 0.1V to 1.3V with the step of 0.1V. Figure 11 (a) shows the linear fitting function between the output of the RCU and the input on the guarding ring of the Topmetal II- ASIC.

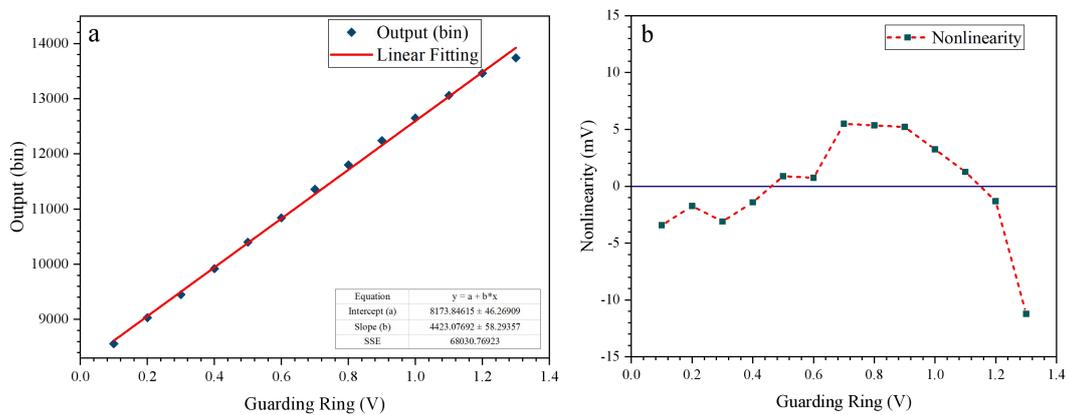

Figure 11. (a) Output as function of input (b) Nonlinearity of each input

Performance of the whole readout system has been tested with the alpha radiation source ($^{241}$Am). Figure 12 shows the test setup, where the FEC is placed in the high vacuum chamber. The residue gas in the vacuum chamber is ionized by the alpha particle and the electrons are collected by the beam monitoring system. Figure 13 shows the electron cluster recorded by the readout system.

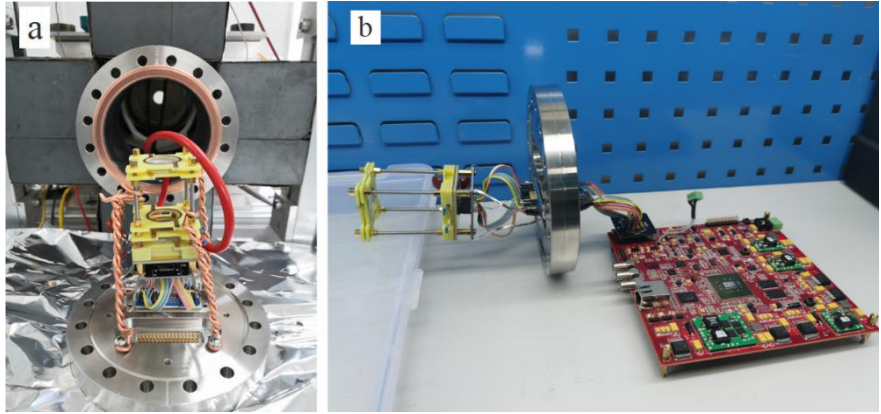

Figure 12. Test setup of the readout system with radiation source

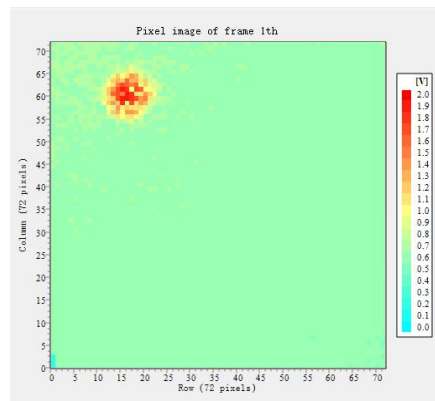

Figure 13. Beam pattern formed by electron hitting Topmetal-II⁻

## 4 Conclusion

This paper discusses the implementation and testing of readout electronics for the new beam monitoring system of the HIRFL-CSR and 320kV platform. The correctness of the test data at different taps numbers shows that the AD9252 data-window is very small. The readout electronics stability was also tested. After 48 hours of continuous work, the nonlinearity of electronics is 0.6%. Testing in a vacuum using an alpha source shows that readout electronics can adapt to different application environments.